\documentclass[10pt, conference, compsocconf]{IEEEtran}

\usepackage{cite}
\usepackage[pdftex]{graphicx}

\usepackage{subfig}
\usepackage{url}
\usepackage{hyperref}

\newcommand{\citep}{\cite}

\begin{document}

\title{User Centered Development of Agent-based Business Process Models and Notations}

\author{
\IEEEauthorblockN{Robert Singer}
\IEEEauthorblockA{
Department of Computer Sciences\\
FH JOANNEUM -- University of Applied Sciences\\
Graz, Austria\\
robert.singer@fh-joanneum.at}
}

\maketitle

\begin{abstract}
We discuss questions about user centric development of business process modeling notations. In the center of our research there is a fully featured multi-enterprise business process platform (ME-BPP) based on the concepts of agent-based business processes, which builds on the formal foundations of the subject-oriented business process management methodology (S-BPM). The platform is implemented based on cloud technology using commercial services. Additionally we developed a "block modeling" technique to find a semantically transparent modeling notation which can be used by novice users to model subject-oriented business process (S-BPM) models. As this is ongoing research there are still serious open questions. But, the presented approach breaks with some of the rules of typical process modeling notations and hopefully stimulates innovation. Additionally we want to continue our research towards the enhancement of our modeling approach towards a user centric "syntax and semantic free" modeling technique to develop user and domain specific modeling notations.
\end{abstract}


\IEEEpeerreviewmaketitle

\section{Introduction}
\label{introduction}

Latest developments in business and technology driven new business models foster more than ever the need for mature business process management (BPM) methodologies and corresponding supporting technologies for \emph{distributed} business processes, so called choreographies. Business processes cannot be seen as isolated workflows for administrative purposes, but as a mean to \emph{coordinate a value system} with supply chain partners. Communication is the very nature of a business process choreography -- or, in other words, any choreography is a set of \emph{structured communication patterns}. That means a choreography defines how work is done, taking into account all involved organizations. Distributed execution of a business process means that every process participant may use their own process execution engine. The overall process is then executed by interconnecting multiple engines. The engines may even run on a mobile device.

This demand is reflected in new developments in the domain of BPM, such as \emph{BPM Platform as a Service} (bpmPaaS), \emph{multi-enterprise Business Process Platform} (ME-BPP), \emph{Cloud BPM}, and \emph{Social BPM}. The term bpmPaaS can be defined~\citep{Anonymous:2012tf} as ``\emph{the delivery of BPM platform capabilities as a cloud service by a service provider}''. A ME-BPP is defined~\citep{Anonymous:2012tf} as ``\emph{high-level conceptual model of a multistakeholder environment, where multi-enterprise applications are operated. multi-enterprise applications are those that are purposely built to support the unique requirements for business processes that span more than one business entity or organization. They replace multiple business applications integrated in serial fashion}''.

For a distributed execution of a process, two important prerequisites are needed: a \emph{suitable process modeling} technique and a \emph{flexible communication platform}~\citep{Aitenbichler:2011lr}. As elaborated in the following sections, we have chosen the agent based approach to model a distributed system. To be more specific, we build on the Subject-oriented BPM methodology, as defined in~\citep{Fleischmann:2012va}.

To implement a communication platform, we need an architecture, which includes a graphical business process and\slash or rule modeling capability, a process registry\slash repository to handle the modeling metadata, a process execution and either a state management engine or a rule engine as minimal requirements. To realize a bpmPaaS and\slash or ME-BPP system a cloud infrastructure is needed to model and execute processes which \emph{span across more than one business entity or organization}.

\subsection{Agent-based BPM (AB-BPM)}
\label{agent-basedbpmab-bpm}

As already discussed, for example recently by~\citep{Sanz.2013} or~\citep{Borger:2011ib}, \emph{traditional} BPM and its supporting technology frameworks (we mean business process management -- or better workflow management -- systems (BPMS, WfMS) as supporting layer for process enactment) have some conceptual and practical limitations. Business processes are more than \emph{algorithmic} workflows (input, black box, output); they often may have deep social aspects, as long as human participants are involved. Therefore, a new view on business processes recently has been promoted under the term \emph{Agent-based BPM} (AB-BPM)~\citep{Sinur:2013vw} or \emph{Subject-oriented BPM} (S-BPM)~\citep{Fleischmann:2010gz}.

There is already a long history of the idea of \emph{interacting agents}. The application of the agent concept into the domain of BPM has emerged from the domain of distributed software~\citep{Fleischmann.1994} by Albert Fleischmann, who developed the Subject-oriented BPM (S-BPM) methodology in the early 2000s based on his PASS\footnote{Parallel Activities Specification Scheme}~\citep{Fleischmann:1987ui} language. All language constructs of PASS can be transformed down to pure CCS\footnote{Calculus of Communicating Systems~\citep{Milner:1989tw}}~\citep{Aitenbichler:2011lr}. The S-BPM methodology enhances the process algebra languages by graphical representations and adds some technical feature definitions. 

\subsection{Distributed BPM}
\label{distributedbpm}

Any collaboration contains more than one subject, so it is per definition a multi agent system, which is a subclass of concurrent systems~\citep{Wooldridge:2009uma} -- which is an important fact as it has consequences for a possible technical implementation. 

The problem of synchronizing multiple processes is not trivial and has been widely studied through the 1970s and 1980s~\citep{BenAri:2006vx}. To understand the problem~\citep{Wooldridge:2009uma}, it is helpful to first consider the way that communication is treated in the object-oriented programming paradigm, that is, communication as a method invocation. The crucial point is, that an object does not have control over the execution of its own public methods -- any other object can execute the object's public methods whenever they want.

\subsection{Contribution}
\label{contribution}

Recently we presented an implementation of a ME-BPP based on our \emph{Structured Information and Communication Technology} (StrICT\footnote{www.strict-solutions.com}) framework, which is based on the S-BPM methodology; the technical aspect have been discussed in~\citep{Rass:2013kv} and~\citep{Singer:Vdg3ngjF}.

In this paper we will focus on the modeling aspects of business processes in general with a concrete application on modeling of agent-based respectively subject-oriented BPM. Further on we will discuss the concept of a ``\emph{semantic and syntax free}'' modeling approach we developed as a generalization of our findings; this is ongoing work, including the development of a prototype implementation using a touch sensitive interface to create models and to create user based modeling languages. Finally we will discuss further possible research directions.

\section{How to Model}
\label{howtomodel}

In this chapter we will work out requirements for and pitfalls of modeling languages. This discussion will show, that it is not sufficient to precisely define the \emph{semantics}, but also the \emph{syntax} of a modeling language. We will also argue, that the predominantly discussed BPMN 2.0 language does not fulfill these requirements to create ``good'' models.

\subsection{The Notion of Model}
\label{thenotionofmodel}

Wand and Weber~\citep{Wand:1990gx} start their seminal work ``\emph{An Ontological Model of an Information System}'' with the following -- and still valid -- remarks:

\begin{quote}

``The computer science (CS) and information systems (IS) fields are replete with fundamental concepts that are poorly defined.''
\end{quote}

A similar discussion about this topic -- for example -- can be found in ``\emph{The FRISCO Report}''~\citep{Falkenberg:1998vl}. Based on these and other discussions, we think it is very fruitful to start the discussion with the notion of ontology. Doing this we can try to build a better understanding of a \emph{systemic view} on organizations; this leads us to a better understanding of the notion of model in general and of process models in particularly.

Business processes offer a \emph{dynamic view} on organizations as they describe the \emph{states} and \emph{state transitions} of a system or the corresponding world. This also defines the ontological model of a system, as elaborated by Dietz~\citep{Dietz.2006}: ``\emph{The ontological model of a world consists of the specification of its state space and its transition space}''. 

\begin{figure}[htbp]
\centering
\includegraphics[keepaspectratio,width=2in,height=0.75\textheight]{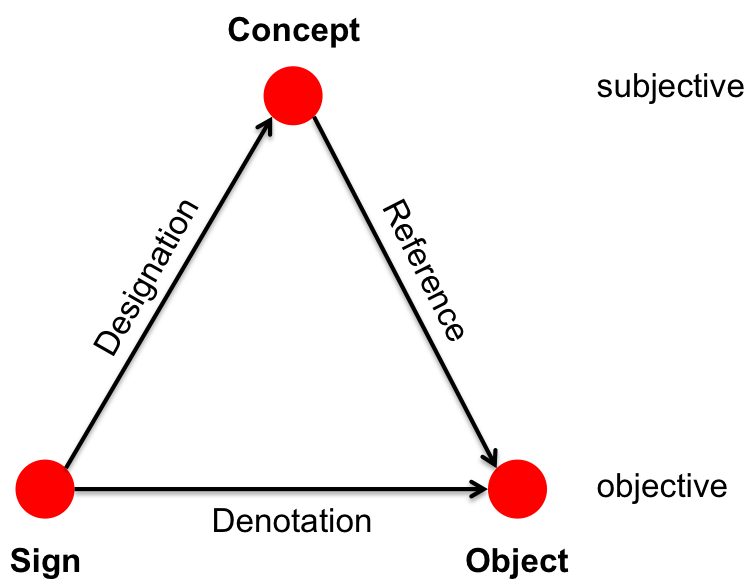}
\caption{The meaning triangle.~\citep{Dietz.2006}}
\label{meaningtriangle}
\end{figure}

If we now want to model a world, we have to understand the meaning triangle from \emph{semiotics} as shown in \autoref{meaningtriangle}.

A \emph{sign} is an object that is used as a representation of something else and is used to \emph{communicate} the \emph{concepts} in our mind. A well-known class of signs are the symbolic signs (structures put into physical substrates). An \emph{object} is an observable and identifiable individual thing; objects are concrete or abstract. A \emph{concept} is a subjective individual thing (a thought in our mind). \emph{Designation} and \emph{denotation} are relevant when we want to communicate. 

A precise formal definition of the construction of a system can be found for example in~\citep{Bunge.1977}, which can be described as: something is a system if it has the following properties: 

\begin{itemize}
\item \emph{Composition}: a set of elements of some category (physical, social{\ldots})

\item \emph{Environment}: a set of elements of the same category; the composition and the environment are disjoint

\item \emph{Structure}: a set of influence bonds among the elements in the composition, and between them and the elements in the environment.

\end{itemize}

\begin{figure}[htbp]
\centering
\includegraphics[keepaspectratio,width=3in,height=0.75\textheight]{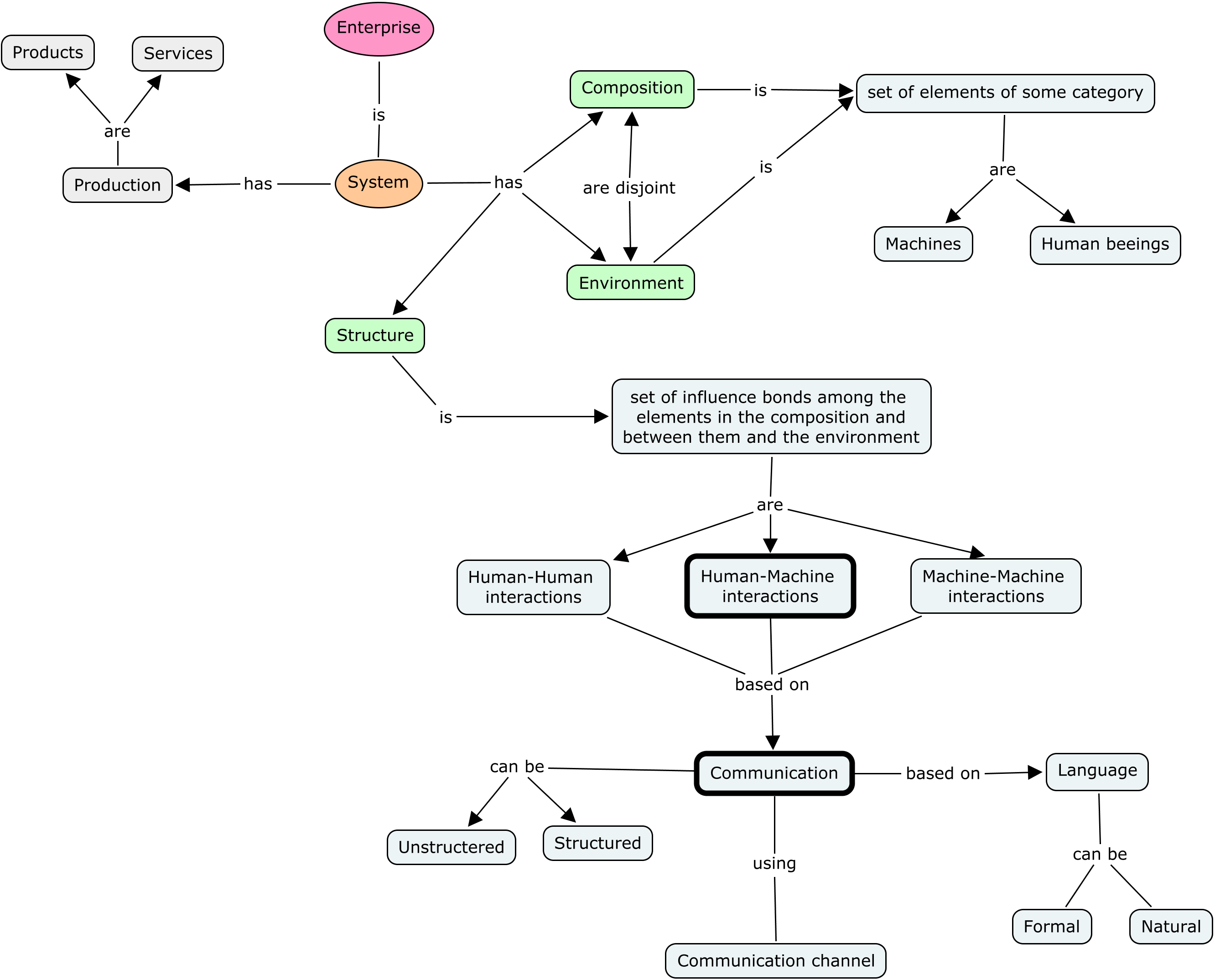}
\caption{The concept of an ontological system view on organizations.}
\label{s-bpm_ontology}
\end{figure}

Dietz further on adds the notion of production~\citep{Dietz.2006}:

\begin{itemize}
\item \emph{Production}: the elements in the composition produce things (goods or services) that are delivered to the elements in the environment. Now, an organization or company is a collection of \emph{socially} linked human beings; an overview about the different concepts mentioned is depicted in \autoref{s-bpm_ontology}.

\end{itemize}

\begin{figure}[htbp]
\centering
\includegraphics[keepaspectratio,width=3in,height=0.75\textheight]{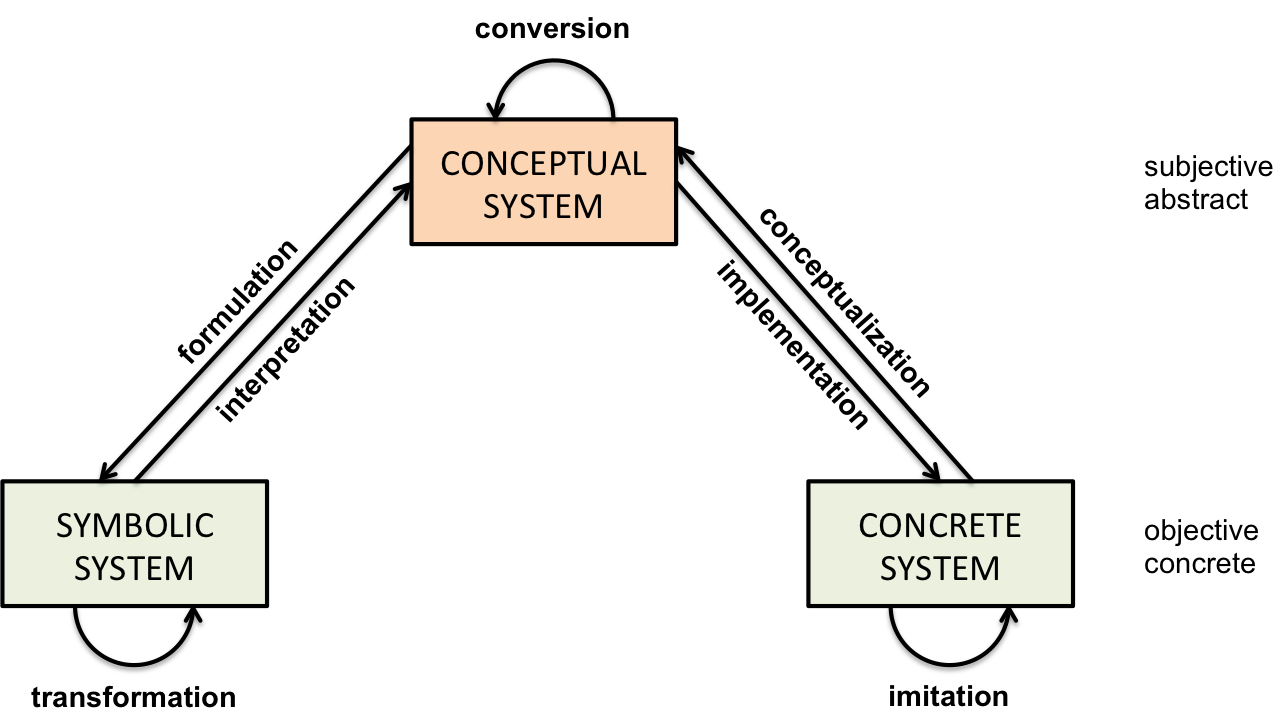}
\caption{The model triangle.~\citep{Dietz.2006}}
\label{modeltriangle}
\end{figure}

Following the argumentation of Dietz~\citep{Dietz.2006}, three gross categories of systems can be distinguished: concrete systems, symbolic systems, and conceptual systems as depicted in \autoref{modeltriangle}.

The conceptual model of a concrete system is called a \emph{conceptualization}; for example, a business process model is a conceptualization of the business processes of an organization or firm. A concrete model of a conceptual system is called an \emph{implementation}; for example an enacted business process is an implementation of a business process model. A conceptual model of a conceptual system is called a \emph{conversion}. A symbolic model of a conceptual system is called a \emph{formulation}; a symbolic system is expressed in formal language -- the notation to represent the model. A conceptual model of a symbolic system is called an \emph{interpretation}. A symbolic model of a symbolic system is called a \emph{transformation}. 

What is the conclusion? Firstly, it is important that the \emph{imitation} of a \emph{concrete} system is never the same. Secondly, concepts are only in our minds and therefore subjective. Thirdly, as a consequence all stages include \emph{social} interaction between human beings to \emph{construct a socially accepted} view of the concrete system. 

Finally, it seems to be clear that it would be a natural approach to use the same language to define business processes as to describe an ontology as a model of a system or organization; \emph{id est} the \emph{state} and \emph{transition space} (ontology view) of \emph{communicating agents} (system view).

\subsection{Requirements for Notations}
\label{requirementsfornotations}

Diagrams can convey information more precisely than ordinary language~\citep{Bertin:1983fe}~\citep{Larkin:1987cq}. As discussed in~\citep{Moody:2009ei} the human mind has separate systems for processing pictorial and verbal material -- according to \emph{dual channel theory}. Visual representations are processed in parallel by the visual system, textual representations are processed serially by the language system~\citep{Bertin:1983fe}. Only diagrammatic presentations are able to show (complex) relations \emph{at once}.

The \emph{anatomy of a visual notation} is worked out very clearly by~\citep{Moody:2009ei}:

\begin{quote}

A visual notation (or visual language, graphical notation, diagramming notation) consists of a set of graphical symbols (\emph{visual vocabulary}), a set of compositional rules (\emph{visual grammar}) and definitions of the meaning of each symbol (\emph{visual semantics}). The visual vocabulary and visual grammar together form the \emph{visual} (or \emph{concrete}) \emph{syntax}. Graphical symbols are used to \emph{symbolize} (perceptually represent) \emph{semantic constructs}, typically defined by a \emph{metamodel}. The meanings of graphical symbols are defined by mapping them to the constructs they represent. A valid expression in a visual notation is called a \emph{visual sentence} or \emph{diagram}. Diagrams are composed of \emph{symbol instances} (\emph{tokens}), arranged according to the rules of the visual grammar.
\end{quote}

But, just presenting information in a graphical form does not guarantee that it will be worth \emph{a thousand of words}~\citep{Cheng:2001vh}. Most effort is spent on designing semantics, with visual syntax often an afterthought~\citep{Moody:2009ei}. For example, UML does not provide design rationale for any of its graphical conventions~\citep{Moody:2009ei}.

\begin{figure}[htbp]
\centering
\includegraphics[keepaspectratio,width=3in,height=0.75\textheight]{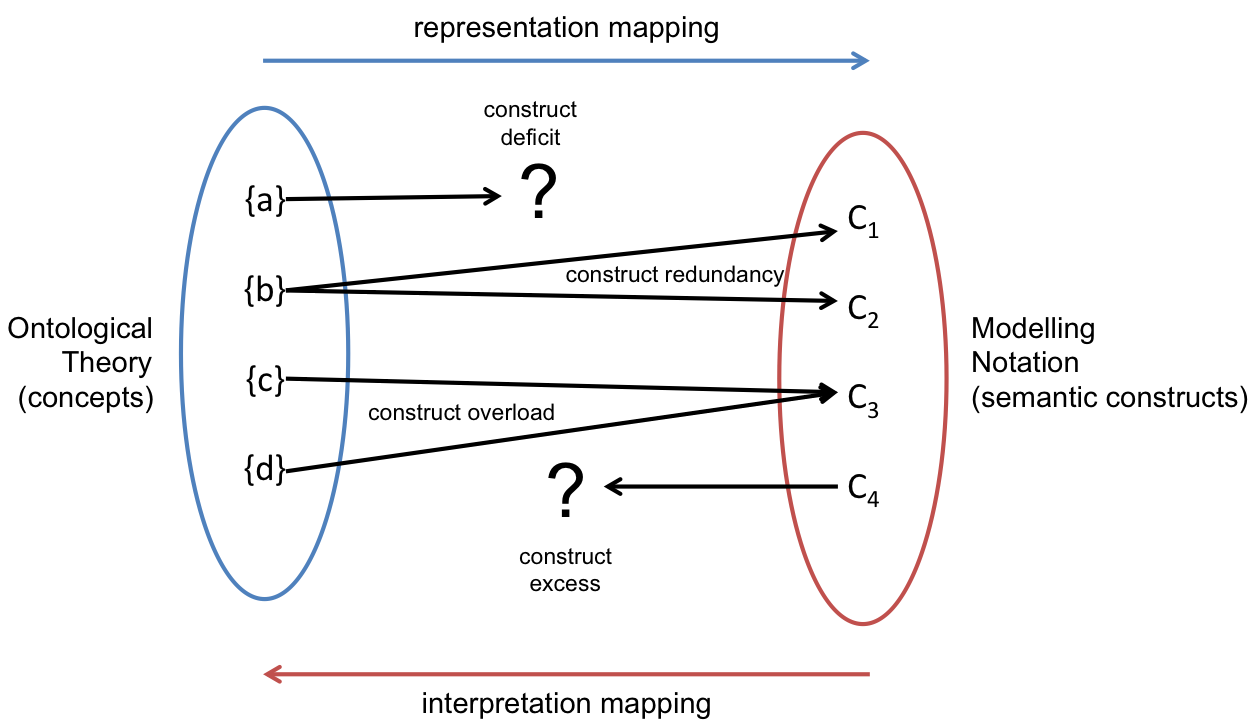}
\caption{Ontological analysis. There should be a 1:1 mapping between ontological concepts and notational constructs.~\citep{Moody:2009ei}}
\label{ontanalysis}
\end{figure}

A widely accepted way to evaluate notations is \emph{ontological analysis}; the most used ontology seems to be the \emph{Bunge-Wand-Weber} (BWW) ontology~\citep{Wand:1990gx}. Ontological analysis involves a two-way mapping between a modeling notation and an ontology. The \emph{interpretation mapping} describes the mapping from the notation to the ontology; the \emph{representation mapping} describes the inverse mapping~\citep{Gehlert:2007fe} as depicted in \autoref{ontanalysis}.

If construct deficits exist, the notation is ontologically incomplete; if any of the other three anomalies exist, it is ontologically unclear. The BWW ontology predicts that ontologically clear and complete notations will be more effective. As elaborated in~\citep{Moody:2009ei}, ontological analysis focuses on content rather than form; if two notations have the same semantics but different syntax, ontological analysis cannot distinguish between them. Moody~\citep{Moody:2009ei} has developed a promising foundation to analyze the \emph{syntactic} aspects of notations in a similar stringent way based on scientific foundations. 

There is a set of principles based on two core concepts~\citep{Moody:2009ei}, we will summarize here as input for further discussions. The developed theory is a so called Type IV theory~\citep{Gregor:2006dk}: a theory for explaining and predicting (how and why). At the top level there is the well accepted theory of communication~\citep{Shannon.1963} and its application to the domain of visual notations:

\begin{quote}

{\ldots}, a diagram creator (sender) encodes information (message) in the form of a diagram (signal) and the diagram user (receiver) decodes this signal. The diagram is encoded using a visual notation (code), which defines a set of conventions that both sender and receiver understand. The medium (channel) is the physical form in which the diagram is presented (e.g., paper, whiteboard, and computer screen). Noise represents random variation in the signal which can interfere with communication. The effectiveness of communication is measured by the match between the intended message and the received message (information transmitted).
\end{quote}

\begin{figure}[htbp]
\centering
\includegraphics[keepaspectratio,width=3.5in,height=0.75\textheight]{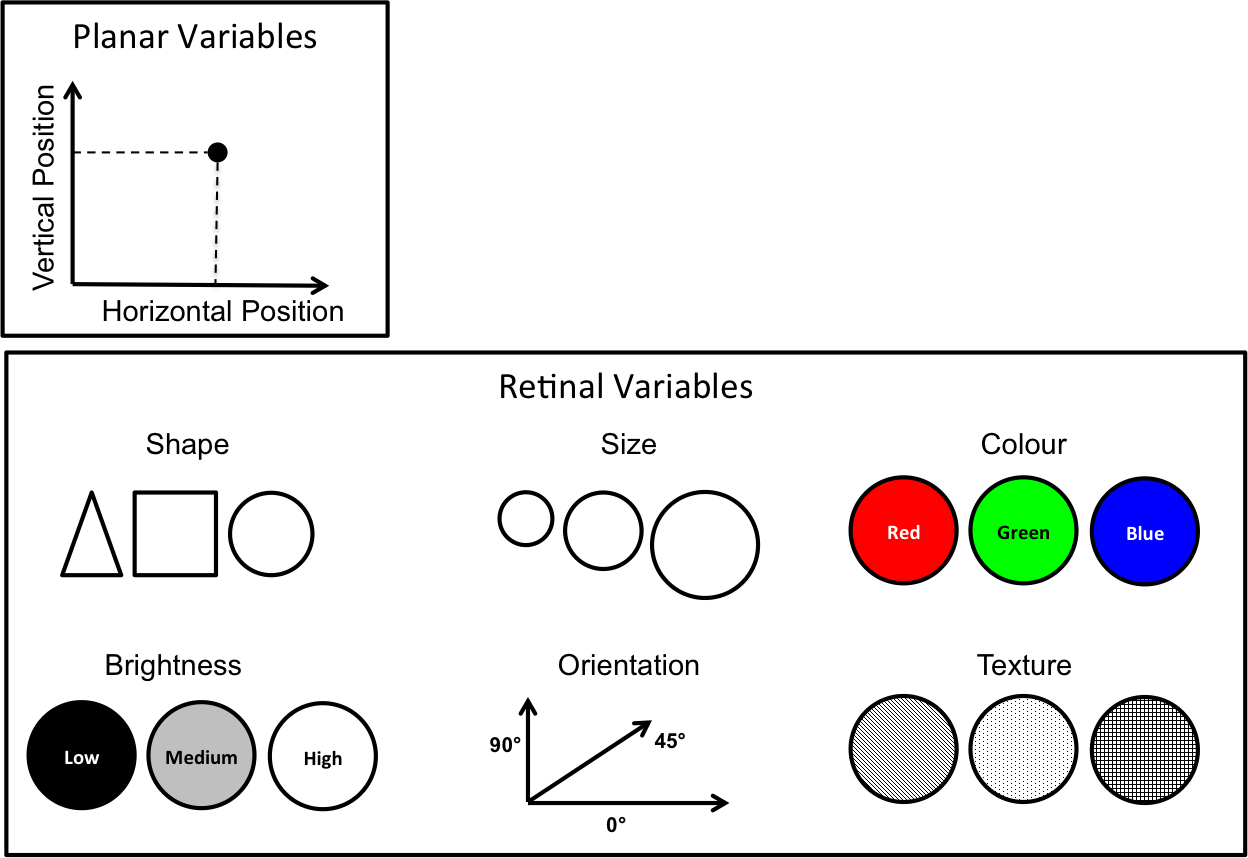}
\caption{Visual variables.~\citep{Bertin:1983fe}; adapted from~\citep{Moody:2009ei}}
\label{bertin}
\end{figure}

Bertin~\citep{Bertin:1983fe} identified eight visual variables that can be used to graphically encode information as depicted in \autoref{bertin}. The decoding side is based on the human decoding processes, which can be divided in two phases: perceptual processing (seeing) and cognitive processing (understanding). As the perceptional processing system is much faster, it is more effective to move as much of the decoding work from the cognitive to it.

Now, based on these theories and empirical evidence Moody has developed a prescriptive theory for visual notations~\citep{Moody:2009ei}, which is formulated as nine principles for designing cognitively effective visual notations, summarized as follows:

\begin{itemize}
\item Semiotic clarity: there should be a 1:1 correspondence between semantic constructs and graphical symbols

\item Perceptual discriminability: different symbols should be clearly distinguishable from each other

\item Semantic transparency: use visual representations whose appearance suggests their meaning

\item Complexity management: include explicit mechanisms for dealing with complexity

\item Cognitive integration: include explicit mechanisms to support integration of information from different diagrams

\item Visual expressiveness: use the full range and capacities of visual variables

\item Dual coding: use text to complement graphics

\item Graphic economy: the number of different graphical symbols should be cognitively manageable

\item Cognitive fit: use different visual dialects for different tasks and audience

\end{itemize}

The method of ontological analysis and the set of principles for designing cognitive effective visual notations, together with the understanding of the notion of model and semiotics assembles a full set of building blocks for a coherent and solid foundation for business process modeling notations. Finally, combining it with a corresponding formal model for business process execution leads to a full theory of business process.

\subsection{Business Process Modeling}
\label{businessprocessmodeling}

As there is (yet) no coherent and \emph{general accepted} theory of business processes and business process management (BPM)~\citep{Borger:2011ib}~\citep{Singer:2011so}~\citep{Sanz.2013}, any way to define process models is the right one; for this purpose domain specific languages (e.g. \emph{notations} such as BPMN or UML) are defined and it cannot be denied that most of them are rooted in the information systems domain. This is a consequence of the fact that information systems engineers need formally defined models without any \emph{semantic} ambiguity. Additionally, modeling is typically conducted by experts, i.e. business analysts and\slash or requirements engineers. But studies~\citep{Caire:2013tg} show that end users understand such expert models very poorly. One of the reasons for this is that it is hard for experts to think like novices, a phenomena called \emph{the curse of knowledge}~\citep{Heath:2007wb}. There are well-known differences in how experts and novices process diagrams~\citep{Cheng:2001vh}.

It is good practice to involve ``users'' in the analysis and design of business processes; this also works in developing software systems (e.g. user-centered design) or in developing new products. Why not involving them in the design process of \emph{notations}? Caire at al.~\citep{Caire:2013tg} have done this for example in a research study regarding requirements engineering (RE) notations (i.e. i*).

\begin{figure}[htbp]
\centering
\includegraphics[keepaspectratio,width=3.5in,height=0.75\textheight]{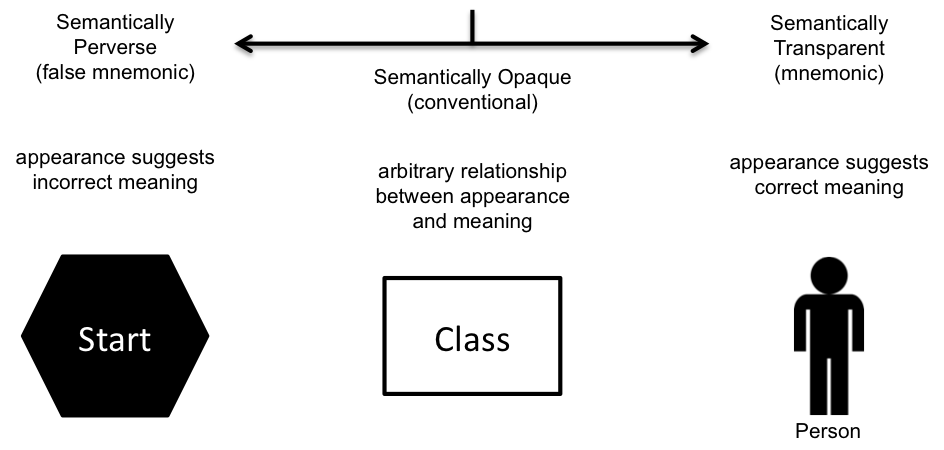}
\caption{Semantic transparency is a continuum.~\citep{Caire:2013tg}}
\label{semtrans}
\end{figure}

The key to designing vial notations that are understandable to na\"{\i}ve users is a property called semantic transparency. This means that the meaning (\emph{semantics}) of a symbol is clear (\emph{transparent}) from its appearance alone; Semantically transparent symbols reduce cognitive load because they have built-in mnemonics~\citep{Petre:1995kt} (see \autoref{semtrans}). 

However, semantic transparency is typically evaluated subjectively: experts (researchers, experts from software vendors) try to estimate the likelihood that novices will be able to infer the meaning of particular symbols. Even when notations are specifically designed for communicating with business stakeholders, members of the target audience are rarely involved. For example, BPMN 2.0 is a notation designed for communicating with business stakeholders, yet no business representatives were involved in the notation design process and no testing was conducted with them prior to its release~\citep{Recker:2010ke}~\citep{Caire:2013tg}.

Business process models are needed to facilitate a shared understanding in the organization; therefore the process creating and documenting the model includes employees unfamiliar with the chosen process design method. Typical workshops on process design employ design tools such as whiteboards, flip charts and post-its to capture knowledge about a current or future process. Informal sketches and diagrammatic drawings were found to be key to any design activity, as they serve as an externalization of one's internal thoughts, and assist in idea creation and problem-solving. 

There is a clear difference how novice and expert modelers conceptualize important domain elements as reported by Wang and Brooks~\citep{WangWang:ij}, who found that novice modelers conceptualize in a fairly linear process in contrast to experts, who have better analysis and critical evaluation skills. Also based on unexperienced modelers, Recker et al.~\citep{Recker:2010fo} developed a range of typical process design archetypes; they found out, that ``\emph{moderate use of graphics and abstract shapes to illustrate a process is more intuitive and would aid the understanding on the concept of process modeling}''. 

\section{Application}
\label{application}

Our focus is on agent-based modeling and execution of business processes. The execution of agent-based business process models has already be discussed in~\citep{Kotremba:2013vz}~\citep{Rass:2013kv}~\citep{Singer:Vdg3ngjF}. The following chapter will discuss the modeling aspects of agent-based business process models based on the S-BPM methodology. Finally we will discuss ongoing work to find \emph{alternative} and more user-friendly ways to develop process models.

\subsection{Subject-oriented BPM}
\label{subject-orientedbpm}

\begin{figure}[htbp]
\centering
\includegraphics[keepaspectratio,width=3.5in,height=0.75\textheight]{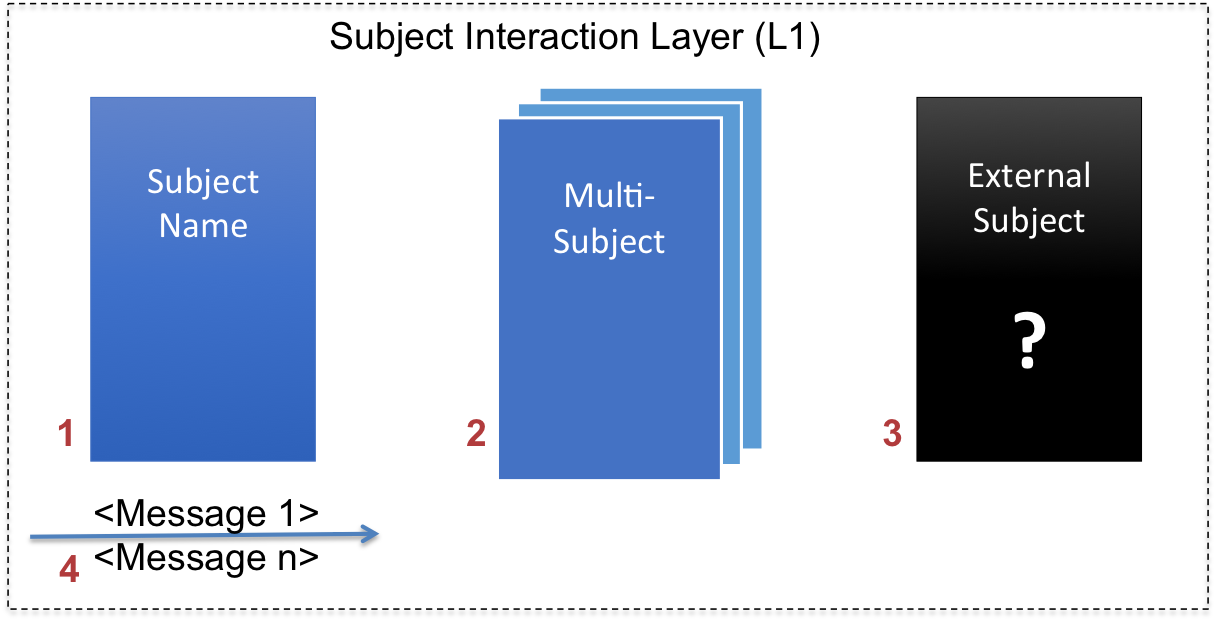}
\caption{Supported S-BPM Language Elements of the \emph{Subject Interaction Diagram} (Layer 1).}
\label{layer1}
\end{figure}

\begin{figure}[htbp]
\centering
\includegraphics[keepaspectratio,width=3in,height=0.75\textheight]{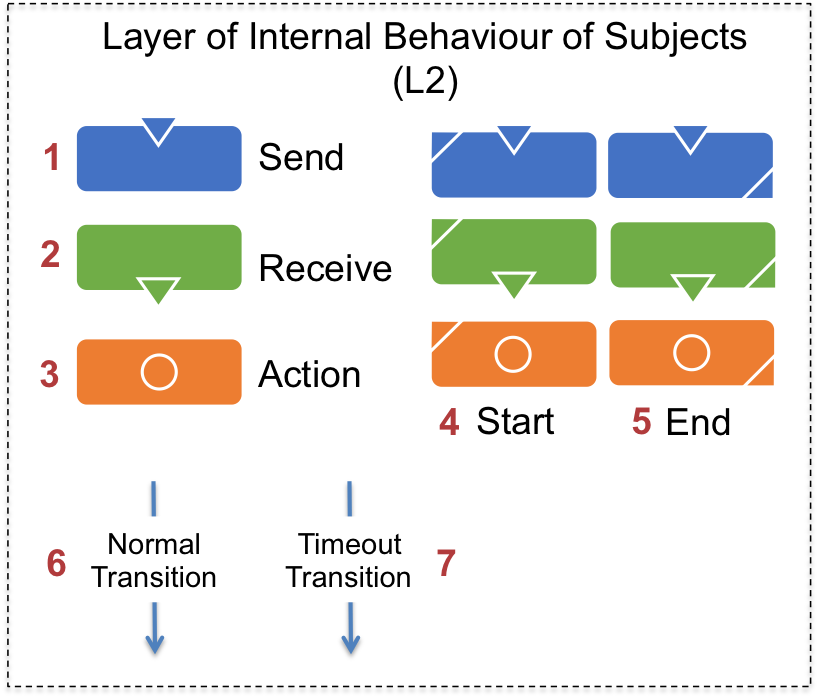}
\caption{Supported S-BPM Language Elements of the \emph{Subject Behavior Diagram} (Layer 2).}
\label{layer2}
\end{figure}

The S-BPM language~\citep{Fleischmann:2012va}, as supported in our execution platform~\citep{Rass:2013kv}~\citep{Singer:Vdg3ngjF}, is depicted in \autoref{layer1} and \autoref{layer2}.

The \emph{Subject Interaction Diagram} (SID) defines the \emph{Subjects} (1) and the unidirectional \emph{Channels} (4) between them. These channels establish the communication between the subjects and enables to send and receive messages at runtime. A \emph{Multi-subject} allows to send a message to more than one agent (an agent is an instance of a subject); an \emph{External-subject} allows to model a subject without knowing the internal behavior, for example another choreography which participants are not part of the own organization.

The \emph{Subject Behavior Diagram} (SBD) defines the internal behavior of a subject; \emph{Send} (1), \emph{Receive} (2) and \emph{Action} (3) are the fundamental activities for this diagram. The internal behavior of subjects has a minimum of one \emph{Start} (4) and one \emph{End} (5) activity. Any activity is marked with a flag to denote it as start or end activity. The normal control flow is defined as explicit \emph{Transition} between activities. \emph{Timeout Transition} are based on a relative time and model exceptional behavior to prevent dead lock situations or service level problems in case of no answer in a defined timeframe.

Contrary to BPMN 2.0, it can be easily seen that this notation has a 1:1 fit between \emph{concepts} and \emph{semantic constructs} (ontological analysis). It also corresponds with the notion of system (communicating agents) and models the state and transition space (ontological model). Nevertheless, at first glance it seems to be much simpler and therefore cognitively more effective; the advantage of easily bridging the gap (from model to execution) is evident, but there seem to be (hidden) mental hampering factors in the field of modeling, as experience shows. One could be, for example, that current software implementations do use two different views as elaborated above. Other cognitive hampering factors have yet to be investigated.

\subsection{Alternative Ways to Model}
\label{alternativewaystomodel}

\begin{figure}[htbp]
\centering
\includegraphics[keepaspectratio,width=3.5in,height=0.75\textheight]{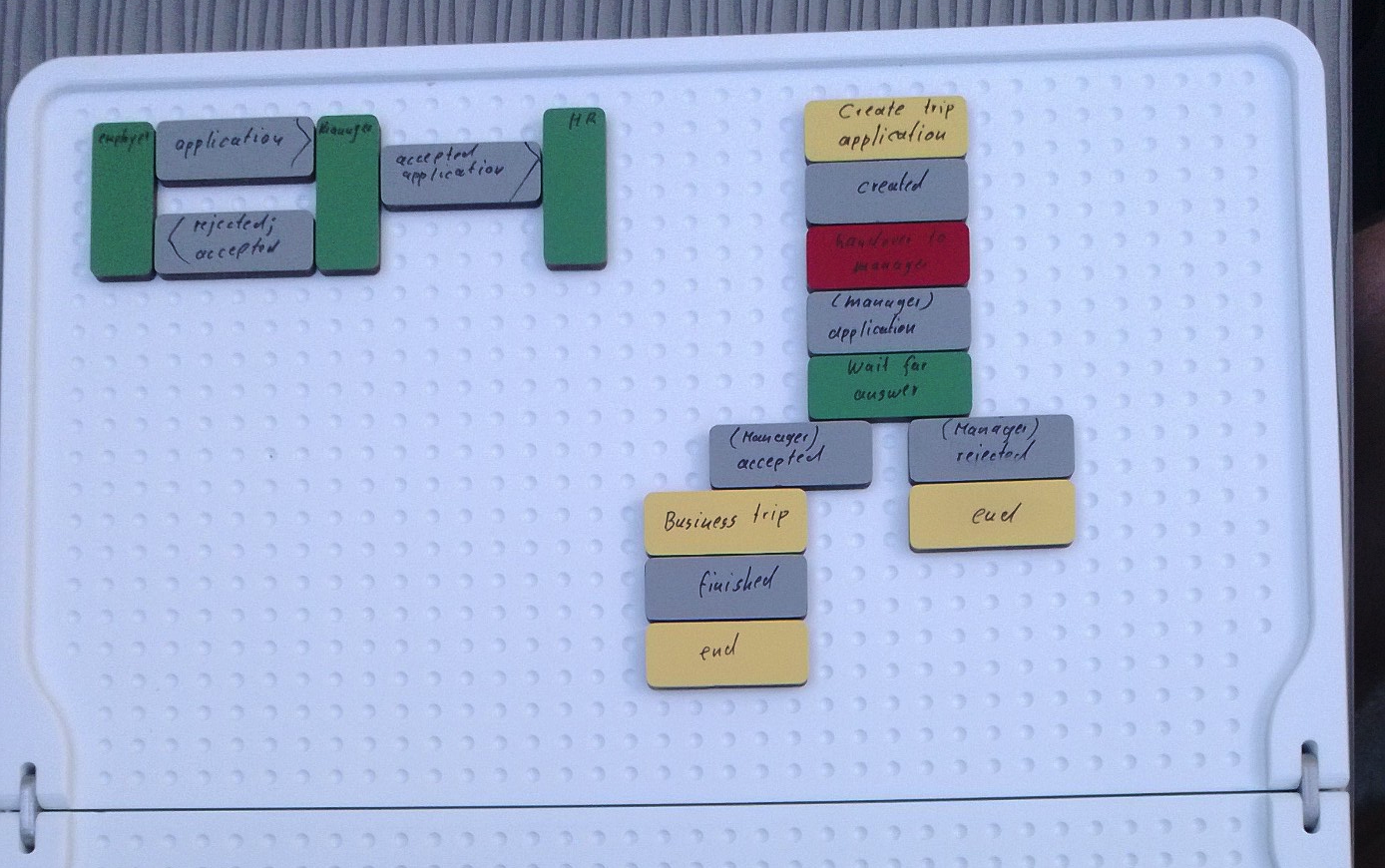}
\caption{S-BPM process model using the \emph{Build book}.}
\label{buildbook}
\end{figure}

Recently Fleischmann ~\citep{Fleischmann:2013uy} has presented a \emph{haptic} way to model S-BPM processes. The main idea is to use \emph{building blocks} to model the processes (\emph{Build Book}), as can be seen in \autoref{buildbook}.

The idea of this and other methods (for example, Metasonic Touch\footnote{http:/\slash www.metasonic.de\slash en\slash touch}) are to \emph{better} involve process participants -- typically untrained for the proposed notation or method -- into the modeling process. People should focus on their work, not on understanding a notation.

Both methods have rather obvious obstacles: for example, they can be used to demonstrate \emph{how} to model, but there are limitations to model \emph{real} business processes: the \emph{Build Book} for example needs a transfer of the model into software for execution using sophisticated picture analysis techniques. Both mentioned methods are limited in their capability to model \emph{hierarchical} and \emph{complex} processes (one of the principles for designing cognitive effective visual notations).

\subsection{Block Modeler}
\label{blockmodeler}

Nevertheless, we think, that the use of blocks offers a very convenient way to model business processes, especially using a modeling notation, such as S-BPM as there are only a very limited number of symbols needed. The core \emph{visual} ideas of our \emph{Block Models} therefore are:

\begin{itemize}
\item all symbols are rectangles (blocks)

\item the semantic and syntax is defined using different colors

\item the blocks are laid out on a canvas

\item the blocks are directly connected on one side; the flow direction is defined by convention (top-down, left-right or vice versa)

\item to add additional needed directed connections, arrows can be used

\end{itemize}

The core concept now has to be transferred onto a technology platform; we propose to do the modeling process on a touch device\footnote{Our modeling prototype has been developed on a Microsoft Surface Pro 2.}. The technology platform should be as flexible as possible, as there are many unclear requirements yet.

\subsubsection{Actual S-BPM Functionality}
\label{actuals-bpmfunctionality}

In our application, we utile the screen of a tablet device and allow the user to drag and drop items from a library section to a stage section. Manipulation includes moving one or several items, selecting one or several items, connecting items, deleting one or several items, and increasing\slash decreasing item width as well as some convenience functionality if dealing with large process models. This includes zooming, moving and an automatically resizing of the stage.

Items are able to change their color or to get a background image (icon). Furthermore they can be labeled with text and are capable of holding properties, which are implemented as a list of \emph{key value pairs}.

\begin{figure}[htbp]
\centering
\includegraphics[keepaspectratio,width=3in,height=0.75\textheight]{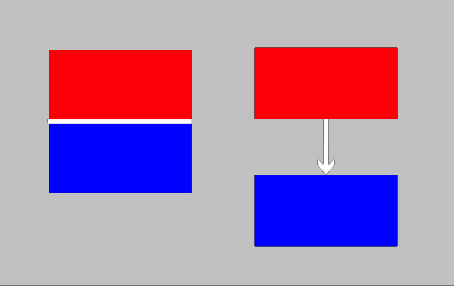}
\caption{Connection possibilities of two elements.}
\label{blockmodel_1}
\end{figure}

There are two ways of connecting items with each other. One possibility is the implicit method by concatenating blocks as described. Another possibility to create a connection between items is to use arrows. This is necessary when the flow is not sequential and needs to skip certain elements. How items are connected does not change their behavior, to illustrate this \autoref{blockmodel_1} shows both ways of connecting items.

\begin{figure}[htbp]
\centering
\includegraphics[keepaspectratio,width=3in,height=0.75\textheight]{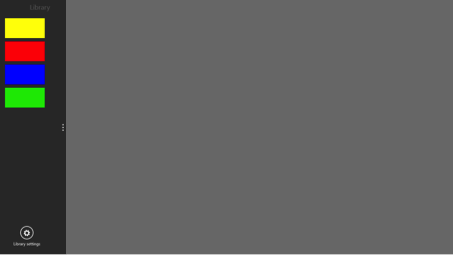}
\caption{The Block Modeler app.}
\label{blockmodel_2}
\end{figure}

The editor app is depicted in \autoref{blockmodel_2}. Two shades of gray indicate the two main areas of the editor. Dark gray represents the library with the available symbols. The lighter shade of gray which spans the biggest area in the application is what is internally named the stage. The stage is an UI element of type canvas, a canvas allows to have elements which can be arbitrary placed on it. The canvas can be zoomed in and elements can be moved around.

The intention of the modeling app is the need to have a flexible tool for further research and elaboration of the block modeling methodology. Anyway, we have to consider, that a \emph{fully functional} modeling tool leads to many questions to be answered -- especially if we want to find \emph{better} ways to involve users in the modeling process. 

The choice of platform allows to design a very interactive behavior with strong visual feedback mechanisms (in further development steps we also plan to implement the simulation of ``physical'' effects). For example, there is a ``docking'' mechanism, so overlapping blocks ``jump'' when released into the nearest valid position forming a connection between the blocks; alternatively it is also allowed to position blocks anywhere on the canvas (e.g. to have more than one model on it). It is also convenient to move models or parts of models around on the canvas.

\begin{figure}[htbp]
\centering
\includegraphics[keepaspectratio,width=3in,height=0.75\textheight]{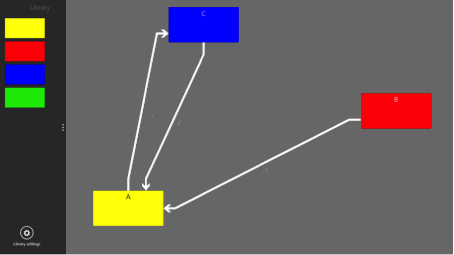}
\caption{A model with arbitrary positioned and connected blocks.}
\label{connectedblocks}
\end{figure}

As we could not find a ready to use library with classes to draw orthogonal directed arcs between blocks, we has to implement some \emph{basic} algorithms for this, as depicted in \autoref{connectedblocks}.

\begin{figure}[htbp]
\centering
\includegraphics[keepaspectratio,width=2.5in,height=0.75\textheight]{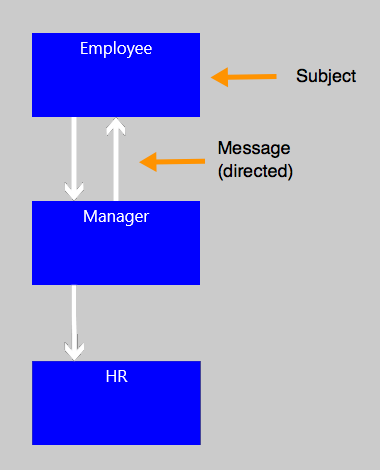}
\caption{Subject model as block model.}
\label{subject_block}
\end{figure}

\begin{figure}[htbp]
\centering
\includegraphics[keepaspectratio,width=3in,height=0.75\textheight]{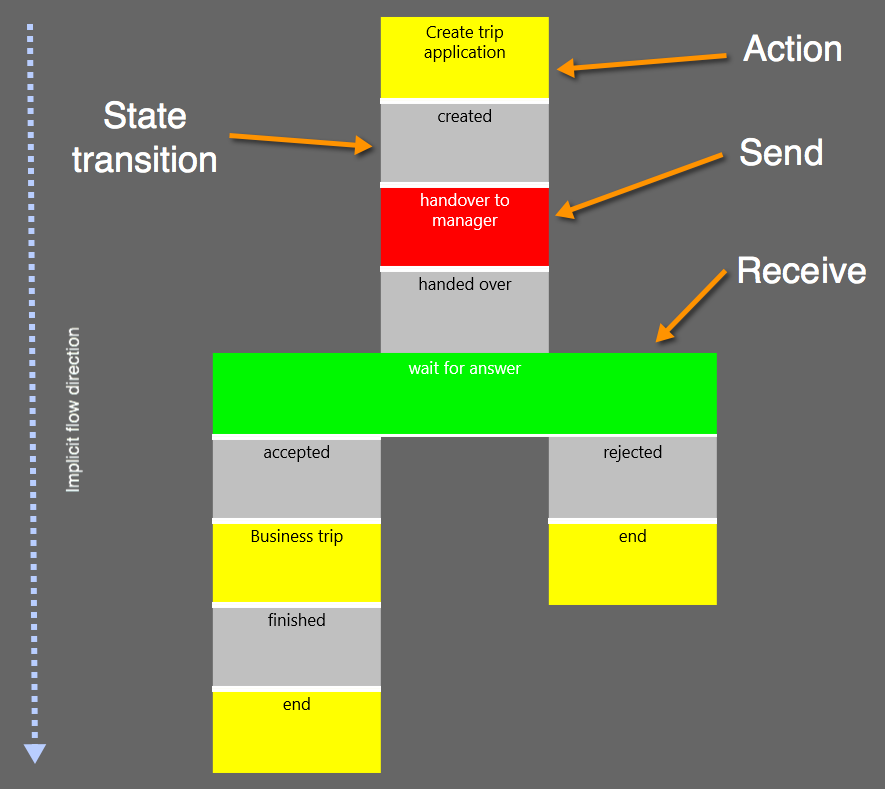}
\caption{Internal behavior as block model.}
\label{behaviour_block}
\end{figure}

It is now possible to create a S-BPM process in analogy to the process shown in \autoref{buildbook}. The subject interaction view is depicted in \autoref{subject_block}; we implemented the concept of hierarchical modeling, so it is possible to drill down into the internal behavior of a subject; this is depicted in \autoref{behaviour_block}. Messages between subjects can actually be added as key-value pairs. After that, the process can be persisted as XML-file and uploaded for execution.

There is one main problem: how to define the direction of flow or relations? We need a well defined and transparent syntax to reflect this -- without using conventions (from left to right, for example). Finally, all models need to be persisted and it must be clear from the graphical representation which elements are \emph{connected}, including a possible \emph{direction} (typically visualized as \emph{directed arc}).

\subsubsection{Semantic- and Syntax Free Modling}
\label{semantic-andsyntaxfreemodling}

An interesting point now is, that the concept of block modeling can be generalized towards a -- as we call it -- \emph{semantic and syntax free} modeling method. The idea is as follows: people involved in \emph{any} modeling problem could use the block modeling method to \emph{define} a problem or domain specific semantic and syntax. Afterwards they could use this defined semantic and syntax to create a domain specific model. Using the concepts of ontology and the principles for designing cognitive effective visual notations this could improve the capabilities of people for conceptualization; the use of a touch screen further enables collaborative and interactive work (in small groups). In this view, S-BPM is only one possible language which can be used to create \emph{block models}. 

\section{Discussion}
\label{discussion}

Latest research proved that symbols designed by na\"{\i}ve subjects increase comprehensibility by a factor of almost 4 and reduce interpretation errors by more than 80\% over notations designed in the traditional way. The average semantic transparency of novice-generated symbols was more than 5 times that of expert-generated symbols. Further on, semantic transparency significantly increases recognition accuracy and reduces interpretation errors.~\citep{Caire:2013tg}

The most difficult attribute of business process models is the fact, that we try to visualize dynamic behavior using static visualizations; another one is to use a plethora of different symbols. The concept of block models opens some interesting questions with the intention to better understand the way how people, novice to modeling in general and especially inexperienced to model semantic precise business process models, can improve their related capabilities.

To this date there are some serious open questions, but in our opinion the block modeling methodology offers a promising research direction to understand cognitive hampering factors to understand process models.

The idea of a \emph{syntax and semantic free} editor further on leads to the following research questions we will tackle in the future:

\begin{itemize}
\item Is it possible and useful to enhance the syntax according the full set of visual variables~\citep{Bertin:1983fe}?

\item How to \emph{store} the semantic and syntax in the modeler, i.e. definition of the modeling language (conformance check during modeling etc.)?

\item Is it possible to find a cognitive easy method to visualize flow and relations (beside the typical use of directed arcs)?

\end{itemize}

These research has to go along with studies of modeling in the field; i.e. to support unexperienced people to learn how to conceptualize in general and how to create business process models. This is supported by the use of a tool, which can be adapted as learning increases. 

Another interesting question could be, if it is possible to \emph{model instances} instead of \emph{process models}, which include all \emph{possible} instances; a software could merge all possible instances into a \emph{general model}, a \emph{blue print} of a business process. This is only possible using notations which are mathematically precise, as the S-BPM method.

\bibliographystyle{IEEEtran}
\bibliography{AS-BPM-2014}

\begin{thebibliography}{10}
\providecommand{\url}[1]{#1}
\csname url@samestyle\endcsname
\providecommand{\newblock}{\relax}
\providecommand{\bibinfo}[2]{#2}
\providecommand{\BIBentrySTDinterwordspacing}{\spaceskip=0pt\relax}
\providecommand{\BIBentryALTinterwordstretchfactor}{4}
\providecommand{\BIBentryALTinterwordspacing}{\spaceskip=\fontdimen2\font plus
\BIBentryALTinterwordstretchfactor\fontdimen3\font minus
  \fontdimen4\font\relax}
\providecommand{\BIBforeignlanguage}[2]{{%
\expandafter\ifx\csname l@#1\endcsname\relax
\typeout{** WARNING: IEEEtran.bst: No hyphenation pattern has been}%
\typeout{** loaded for the language `#1'. Using the pattern for}%
\typeout{** the default language instead.}%
\else
\language=\csname l@#1\endcsname
\fi
#2}}
\providecommand{\BIBdecl}{\relax}
\BIBdecl

\bibitem{Anonymous:2012tf}
``{Gartner BPM Hype Cycle},'' Gartner, Tech. Rep., 2012.

\bibitem{Aitenbichler:2011lr}
E.~Aitenbichler, S.~Borgert, and M.~M{\"u}hlh{\"a}user, ``{Distributed
  Execution of S-BPM Business Processes},'' in \emph{S-BPM ONE 2010}.\hskip 1em
  plus 0.5em minus 0.4em\relax Berlin, Heidelberg: Springer, 2011, pp. 19--35.

\bibitem{Fleischmann:2012va}
A.~Fleischmann, W.~Schmidt, C.~Stary, S.~Obermeier, and E.~B{\"o}rger,
  \emph{{Subject-Oriented Business Process Management}}.\hskip 1em plus 0.5em
  minus 0.4em\relax Springer, 2012.

\bibitem{Sanz.2013}
\emph{{Enabling Customer Experience and Front-Office Transformation through
  Business Process Engineering}}, IEEE Conference on Business
  Informatics.\hskip 1em plus 0.5em minus 0.4em\relax Vienna: Key Note, 2013.

\bibitem{Borger:2011ib}
E.~B{\"o}rger, ``{Approaches to modeling business processes: a critical
  analysis of BPMN, workflow patterns and YAWL},'' \emph{Software {\&} Systems
  Modeling}, 2011.

\bibitem{Sinur:2013vw}
J.~Sinur, J.~Odell, and P.~Fingar, \emph{{Business Process Management: The Next
  Wave}}, ser. Harnessing Complexity with Intelligent Agents.\hskip 1em plus
  0.5em minus 0.4em\relax Meghan-Kiffer Press, 2013.

\bibitem{Fleischmann:2010gz}
A.~Fleischmann, ``{What is S-BPM?}'' in \emph{S-BPM{\textasciitilde}ONE -
  Setting the Stage for Subject-Oriented Business Process Management}.\hskip
  1em plus 0.5em minus 0.4em\relax Springer, 2010, pp. 85--106.

\bibitem{Fleischmann.1994}
------, \emph{{Distributed Systems: Software design and implementation}}.\hskip
  1em plus 0.5em minus 0.4em\relax Springer, 1994.

\bibitem{Fleischmann:1987ui}
------, \emph{{PASS - A Technique for Specifying Communication
  Protocols}}.\hskip 1em plus 0.5em minus 0.4em\relax North-Holland Publishing
  Co., May 1987.

\bibitem{Milner:1989tw}
R.~Milner, \emph{{Communication and Concurrency}}.\hskip 1em plus 0.5em minus
  0.4em\relax Prentice-Hall, 1989.

\bibitem{Wooldridge:2009uma}
M.~Wooldridge, \emph{{An Introduction to MultiAgent Systems}}.\hskip 1em plus
  0.5em minus 0.4em\relax John Wiley {\&} Sons, Jun. 2009.

\bibitem{BenAri:2006vx}
M.~Ben-Ari, \emph{{Principles of Concurrent and Distributed
  Programming}}.\hskip 1em plus 0.5em minus 0.4em\relax Pearson Education,
  2006.

\bibitem{Rass:2013kv}
S.~Ra{\ss}, J.~Kotremba, and R.~Singer, ``{The S-BPM Architecture: A Framework
  for Multi-agent Systems},'' \emph{2013 IEEE/WIC/ACM International Joint
  Conferences on Web Intelligence (WI) and Intelligent Agent Technologies
  (IAT)}, vol.~3, pp. 78--82, 2013.

\bibitem{Singer:Vdg3ngjF}
R.~Singer, J.~Kotremba, and S.~Ra{\ss}, ``{Modeling and Execution of
  Multienterprise Business Processes},'' in \emph{Conference on Business
  Informatics}.

\bibitem{Wand:1990gx}
Y.~Wand and R.~Weber, ``{An ontological model of an information system},''
  \emph{IEEE Transactions on Software Engineering}, vol.~16, no.~11, pp.
  1282--1292, 1990.

\bibitem{Falkenberg:1998vl}
E.~D. Falkenberg, ``{A Framework Of Information System Concepts},'' Tech. Rep.,
  1998.

\bibitem{Dietz.2006}
J.~L.~G. Dietz, \emph{{Enterprise Ontology: Theory and Methodology}}.\hskip 1em
  plus 0.5em minus 0.4em\relax Springer, 2006.

\bibitem{Bunge.1977}
M.~Bunge, \emph{{Ontology: The Furniture of the World}}, ser. Treatise on basic
  philosophy.\hskip 1em plus 0.5em minus 0.4em\relax Reidel, 1977, vol.~3.

\bibitem{Bertin:1983fe}
J.~Bertin, \emph{{Semiology of graphics}}, ser. diagrams, networks, maps.\hskip
  1em plus 0.5em minus 0.4em\relax esri press, 1983.

\bibitem{Larkin:1987cq}
J.~H. Larkin and H.~A. Simon, ``{Why a Diagram is (Sometimes) Worth Ten
  Thousand Words},'' \emph{Cognitive science}, vol.~11, no.~1, pp. 65--100,
  1987.

\bibitem{Moody:2009ei}
D.~Moody, ``{The "Physics" of Notations: Toward a Scientific Basis for
  Constructing Visual Notations in Software Engineering},'' \emph{IEEE
  Transactions on Software Engineering}, vol.~35, no.~6, pp. 756--779, 2009.

\bibitem{Cheng:2001vh}
P.~Cheng, R.~K. Lowe, and M.~Scaife, ``{Cognitive science approaches to
  understanding diagrammatic representations},'' \emph{Artificial Intelligence
  Review}, 2001.

\bibitem{Gehlert:2007fe}
A.~Gehlert and W.~Esswein, ``{Toward a formal research framework for
  ontological analyses},'' \emph{Advanced Engineering Informatics}, vol.~21,
  no.~2, pp. 119--131, Apr. 2007.

\bibitem{Gregor:2006dk}
S.~Gregor, ``{The nature of theory in information systems},'' \emph{MIS
  quarterly}, 2006.

\bibitem{Shannon.1963}
C.~E. Shannon and W.~Weaver, \emph{{The Mathematical Theory of
  Communication}}.\hskip 1em plus 0.5em minus 0.4em\relax Univ. of Illinois
  Press, 1963.

\bibitem{Singer:2011so}
R.~Singer and E.~Zinser, ``{Business Process Management - Do We Need a New
  Research Agenda?}'' in \emph{Subject-Oriented Business Process
  Management}.\hskip 1em plus 0.5em minus 0.4em\relax Springer, 2011, pp.
  220--226.

\bibitem{Caire:2013tg}
P.~Caire, N.~Genon, and P.~Heymans, ``{Visual notation design 2.0: towards user
  comprehensible requirements engineering notations},'' \emph{Requirements
  {\ldots}}, 2013.

\bibitem{Heath:2007wb}
C.~Heath and D.~Heath, \emph{{Made to Stick}}, ser. Why Some Ideas Take Hold
  and Others Come Unstuck.\hskip 1em plus 0.5em minus 0.4em\relax Arrow Books
  Limited, 2007.

\bibitem{Petre:1995kt}
M.~Petre, ``{Why looking isn't always seeing: readership skills and graphical
  programming},'' \emph{Communications of the ACM}, vol.~38, no.~6, pp. 33--44,
  Jun. 1995.

\bibitem{Recker:2010ke}
J.~Recker, ``{Opportunities and constraints: the current struggle with BPMN},''
  \emph{Business Process Management Journal}, vol.~16, no.~1, pp. 181--201,
  2010.

\bibitem{WangWang:ij}
W.~Wang and R.~J. Brooks, ``{Empirical investigations of conceptual modeling
  and the modeling process},'' in \emph{2007 Winter Simulation
  Conference}.\hskip 1em plus 0.5em minus 0.4em\relax IEEE, pp. 762--770.

\bibitem{Recker:2010fo}
J.~Recker, N.~Safrudin, and M.~Rosemann, ``{How Novices Model Business
  Processes},'' vol. 6336, pp. 29--44, 2010.

\bibitem{Kotremba:2013vz}
J.~Kotremba, S.~Ra{\ss}, and R.~Singer, ``{Distributed Business Processes - A
  Framework for Modeling and Execution},'' Sep. 2013.

\bibitem{Fleischmann:2013uy}
C.~Fleischmann, ``{Subject-oriented Process Survey},'' Ph.D. dissertation,
  Vienna University of Technology, Nov. 2013.

\end{thebibliography}

\end{document}